\pdfoutput=1
\documentclass[sigconf, nonacm]{acmart}

\AtBeginDocument{%
  }

\setcopyright{acmcopyright}
\copyrightyear{2022}
\acmYear{2022}
\acmDOI{XXXXXXX.XXXXXXX}

\acmConference[MEMSYS '22]{Proceedings of the International Symposium on Memory Systems}{October--December,
  2022}{Virtual Conference}
\acmPrice{15.00}
\acmISBN{978-1-4503-XXXX-X/18/06}

\usepackage{listings}

\definecolor{darkgreen}{rgb}{0,0.43,0}

\lstdefinelanguage{DRAMml}{
    keywords={},
    otherkeywords={
    ->, ->>, -<>, -o, *\\
    },
    keywordstyle=\color{blue}\bfseries,
    keywords=[2]{Hierarchy, Places, Arcs},
    keywordstyle=[2]\color{darkgreen}\bfseries,
    identifierstyle=\color{black},
    sensitive=false,
    comment=[l]{//},
    morecomment=[s]{/*}{*/},
    commentstyle=\color{gray},
    stringstyle=\color{red},
    morestring=[b]',
    morestring=[b]"
}
\lstdefinelanguage{SVA}{
    keywords={},
    otherkeywords={
    for, begin, end, genvar, generate, logic, property, rose, endproperty, not, assert, posedge, always, if, iff, else
    },
    keywordstyle=\color{blue}\bfseries,
    keywords=[2]{Hierarchy},
    keywordstyle=[2]\color{darkgreen}\bfseries,
    identifierstyle=\color{black},
    sensitive=false,
    comment=[l]{//},
    morecomment=[s]{/*}{*/},
    commentstyle=\color{gray},
    stringstyle=\color{red}
}

\usepackage{tikz}
\usetikzlibrary{patterns,arrows,decorations.pathreplacing}
\usetikzlibrary{arrows.meta}
\usetikzlibrary{positioning}
\usetikzlibrary{positioning,shadows,trees}
\newcommand{\myarcs}{
        \begin{tikzpicture}[baseline={([yshift=-.5ex]current bounding box.center)}]    
            \draw[>=triangle 60,->] (0,0) -- (0.5,0);
        \end{tikzpicture}
}
\newcommand{\myarco}{
        \begin{tikzpicture}[baseline={([yshift=-.5ex]current bounding box.center)}]    
            \draw[-*] (0,0) -- (0.5,0);
        \end{tikzpicture}
}
\newcommand{\myarcd}{
        \begin{tikzpicture}[baseline={([yshift=-.5ex]current bounding box.center)}]    
            \draw[>=triangle 60,->>] (0,0) -- (0.5,0);
        \end{tikzpicture}
}

\newcommand{\myarctj}{
        \begin{tikzpicture}[baseline={([yshift=-1.5ex]current bounding box.center)}]    
            \draw[-{Diamond}, blue] (0,0) --  node [above]{\footnotesize$t_x$} (0.7,0);
        \end{tikzpicture}
}

\usepackage{tikz-timing}[2009/12/09]
\usetikztiminglibrary{overlays}
\tikztimingmetachar{A}{1.0D{\texttt{ACT}}}
\tikztimingmetachar{P}{1.0D{\texttt{PRE}}}
\tikztimingmetachar{X}{1.0D{\texttt{NOP}}}
\tikztimingmetachar{R}{1.0D{\texttt{RD}}}
\tikztimingmetachar{W}{1.0D{\texttt{WR}}}
\tikztimingmetachar{O}{1.0D{\texttt{NOP}}}

\newcommand{\timemeasure}[4]
{
    \draw [red,semithick] ($ (#1) - (-0.1,0) $) -- ($ (#1) - (-0.1,#3) -(0,1) $);
    \draw [red,semithick] ($ (#2) - (-0.1,0) $) -- ($ (#2) - (-0.1,#3) -(0,1) $);
    \draw [red,semithick,>=triangle 60, <->] ($ (#1) - (-0.1,#3) $) -- ($ (#2) -
    (-0.1,#3) $) node [below,midway] {#4};
}

\begin{document}

\title{A Framework for Formal Verification of DRAM Controllers}

\author{Lukas Steiner}
\orcid{0000-0003-2677-6475}
\affiliation{%
  \institution{Microelectronic Systems Design Research Group, TU Kaiserslautern}
  \streetaddress{Erwin-Schrödinger-Straße 12}
  \city{Kaiserslautern}
  \country{Germany}
  \postcode{67663}
}
\email{lsteiner@eit.uni-kl.de}

\author{Chirag Sudarshan}
\affiliation{%
  \institution{Microelectronic Systems Design Research Group, TU Kaiserslautern}
  \streetaddress{Erwin-Schrödinger-Straße 12}
  \city{Kaiserslautern}
  \country{Germany}
  \postcode{67663}
}
\email{sudarshan@eit.uni-kl.de}

\author{Matthias Jung}
\orcid{0000-0003-0036-2143}
\affiliation{%
  \institution{Fraunhofer Institute for Experimental Software Engineering (IESE)}
  \streetaddress{Fraunhofer-Platz 1}
  \city{Kaiserslautern}
  \country{Germany}
  \postcode{67663}
}
\email{matthias.jung@iese.fraunhofer.de}

\author{Dominik Stoffel}
\orcid{0000-0002-8180-9738}
\affiliation{%
  \institution{Chair of Electronic Design Automation, TU Kaiserslautern}
  \streetaddress{Erwin-Schrödinger-Straße 12}
  \city{Kaiserslautern}
  \country{Germany}
  \postcode{67663}
}
\email{stoffel@eit.uni-kl.de}

\author{Norbert Wehn}
\orcid{0000-0002-9010-086X}
\affiliation{%
  \institution{Microelectronic Systems Design Research Group, TU Kaiserslautern}
  \streetaddress{Erwin-Schrödinger-Straße 12}
  \city{Kaiserslautern}
  \country{Germany}
  \postcode{67663}
}
\email{wehn@eit.uni-kl.de}

\renewcommand{\shortauthors}{Steiner et al.}

\begin{abstract}

The large number of recent JEDEC DRAM standard releases and their increasing feature set makes it difficult for designers to rapidly upgrade the memory controller IPs to each new standard.
Especially the hardware verification is challenging due to the higher protocol complexity of standards like DDR5, LPDDR5 or HBM3 in comparison with their predecessors.
With traditional simulation-based verification it is laborious to guarantee the coverage of all possible states, especially for control flow rich memory controllers. 
This has a direct impact on the time-to-market.
A promising alternative is formal verification because it allows to ensure protocol compliance based on mathematical proofs. 
However, with regard to memory controllers no fully-automated verification process has been presented in the state-of-the-art yet, which means there is still a potential risk of human error.
In this paper we present a framework that automatically generates SystemVerilog Assertions for a DRAM protocol. 
In addition, we show how the framework can be used efficiently for different tasks of memory controller development.







\end{abstract}

%


\maketitle
\begin{figure}[t]
    \centering
    \includegraphics[width=1.0\linewidth]{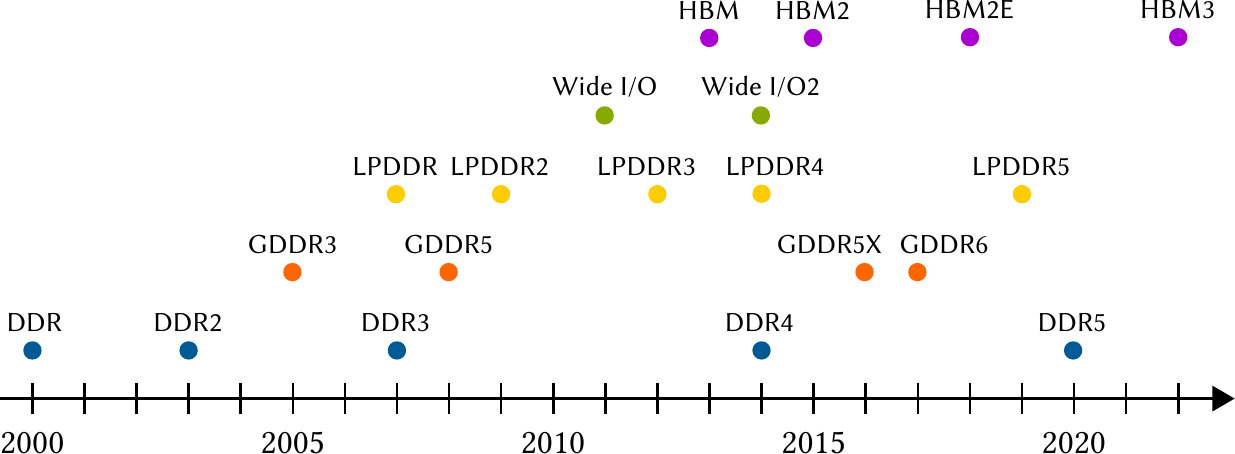}
    \caption{JEDEC Standard Releases Over Time~\cite{junkra_19}}
    \label{fig:jedec_standards}
\end{figure}
\section{Introduction}
Recently, we are witnessing a vast increase in memory-intensive applications (e.g., neural networks), which brings the memory subsystem more into focus.
To keep up with this trend, the DRAM standardization committee JEDEC has released multiple domain-specific standards over the years, shown in Figure~\ref{fig:jedec_standards}, and also upgraded them to meet the constantly increasing bandwidth, storage density and energy efficiency demands of the compute platforms.
This poses a major challenge for system designers because the complex protocols described in new JEDEC standards have to be supported by their memory controller IPs. 
Furthermore, the number of new features introduced in each standard grows substantially from generation to generation, while the time-to-market for today's products is extremely short. 
E.g., the only major upgrade from DDR3 to DDR4 that affected the existing protocol was the introduction of bank groups, while the switch from DDR4 to DDR5 included new refresh commands, different burst lengths, on-die error correction and 3D-stacked devices only to mention some of the new protocol-relevant features.
This is also reflected in the length of each standard (DDR3: 226 pages, DDR4: 270 pages, DDR5: 490 pages) and also projects to the other domains like HBM or LPDDR. 
Moreover, first working samples of DRAM-based processing-in-memory (PIM) devices have been published by Samsung and SK Hynix recently~\cite{kwolee_21,leekim_22}, which require additional extensions on top of the implemented base protocol. 
Due to the complexity of these protocols, the verification of memory controller IPs is extremely challenging nowadays.

Even though simulation-based verification is still widely used in hardware design, it has the major drawback that it cannot prove the absence of errors, but only reveal their presence. 
Especially in control flow rich designs like memory controllers there is a high chance of missing some corner cases with the provided test inputs.
An alternative to simulation-based verification is \textit{formal verification}, which ensures the behavior of a design subject to specified properties based on mathematical proofs. 
This approach has also been applied to DRAM controllers in the past~\cite{datsin_08,kasmic_13}. 
The disadvantage of these works is that all properties have to be extracted manually from the JEDEC documents and translated manually into a property specification language (e.g., \textit{SystemVerilog Assertions} (SVA)), posing a potential source of human error. 
In~\cite{kayabd_14} this problem is addressed by transferring timing diagrams of JEDEC standards into the so-called \textit{Timing Diagram Markup Language} (TDML), from which SVA code is generated automatically. 
Consequently, this approach is limited to those parts of the specification that are illustrated as timing diagrams. However, JEDEC standards also use tables and state machine diagrams to define the protocols.
The authors of \cite{junkra_19} have introduced a comprehensive and formal mathematical model based on \textit{Petri nets} that contains all states, state transitions, commands and timings appearing in today's JEDEC standards. 
This model can also be represented in a compact and machine-readable format with a domain-specific language called \textit{DRAMml}, which serves as a unique formal description. 
As a result, DRAMml can directly be used for the specification of future JEDEC standards.
In their paper, an executable software model for simulation-based verification of virtual controller prototypes is generated from the description.

In this paper, we extend the generation approach to formal verification with the following new contributions:
\begin{itemize}
    \item We present a framework that automatically generates SVA from DRAMml.
    \item Besides ensuring protocol compliance, we show that the framework can also identify the unsupported protocol features of a given controller.
    \item We present a technique to detect timing-related performance limitations of a given controller using slightly-modified properties.
    \item We introduce a methodology to upgrade an existing memory controller to a different JEDEC standard by revealing the minimum set of required changes.
\end{itemize}

In our experiments we show that for one specific configuration the framework detected protocol violations and overestimated timings within an open-source DDR4 controller. In addition, we were able to identify all unimplemented protocol features of the controller without any prior knowledge or documentation. This demonstrates the power and versatility of formal verification for memory controller development. 
 
The remaining paper is structured as follows: Section~\ref{sec:background} introduces the required background. Section~\ref{sec:petri_net} presents a new Petri net model for the DDR4 standard. In Section~\ref{sec:generation} the property generation process within the framework is explained in detail. Experimental results of the framework application are presented in Section~\ref{sec:results}. Section~\ref{sec:related} discusses related work before the paper is concluded in Section~\ref{sec:conclusion}.

\section{Background}\label{sec:background}
In this section we explain the basics of DRAMs, reintroduce timed Petri nets and the DRAM modeling language DRAMml~\cite{junkra_19}, and give an overview of formal verification.

\subsection{DRAM Basics}
DRAM is organized in a multi-hierarchical fashion of channels, ranks, banks, rows and columns. Newer standards add additional hierarchies like bank groups (DDR4), logical ranks (DDR5) or pseudo channels (HBM2). Channels are completely independent entities with separate command/address and data buses. Each channel can be composed of multiple ranks, which are separate devices that also work independently of each other, but share common buses to the controller. Each rank is internally composed of several banks with a shared I/O region. Inside a bank, memory cells are organized in rows and columns. 

Before data can be read (\texttt{RD}) from or written (\texttt{WR}) to cells, a row has to be activated (\texttt{ACT}), and only one row per bank can be active at the same time. In order to switch to another row, the currently active one has to be precharged (\texttt{PRE}) first. Since DRAM is a charge-based memory with leakage effects, each cell has to be refreshed regularly (usually every 64\,ms). The internal refresh operation is triggered externally by the controller with a special refresh command (\texttt{REFA}) and makes the device inaccessible for a certain time. During idle periods, the controller can send DRAM devices into various power-down modes (PDN, SREF) to save power. While in power-down, the devices are also inaccessible. Due to internal delays and shared hardware resources in the devices, a variety of different timings has to be observed by the controller between commands.  

The memory controller itself is composed of a front end and back end. While the front end performs arbitration and scheduling of incoming read and write requests subject to certain priorities, the back end translates the requests into a sequence of commands that complies with the implemented standard. Hence, the back end is also the part of a DRAM controller that has to be adapted to new standards over and over.

\begin{figure*}[t]
    \centering
    \includegraphics[width=1.0\linewidth]{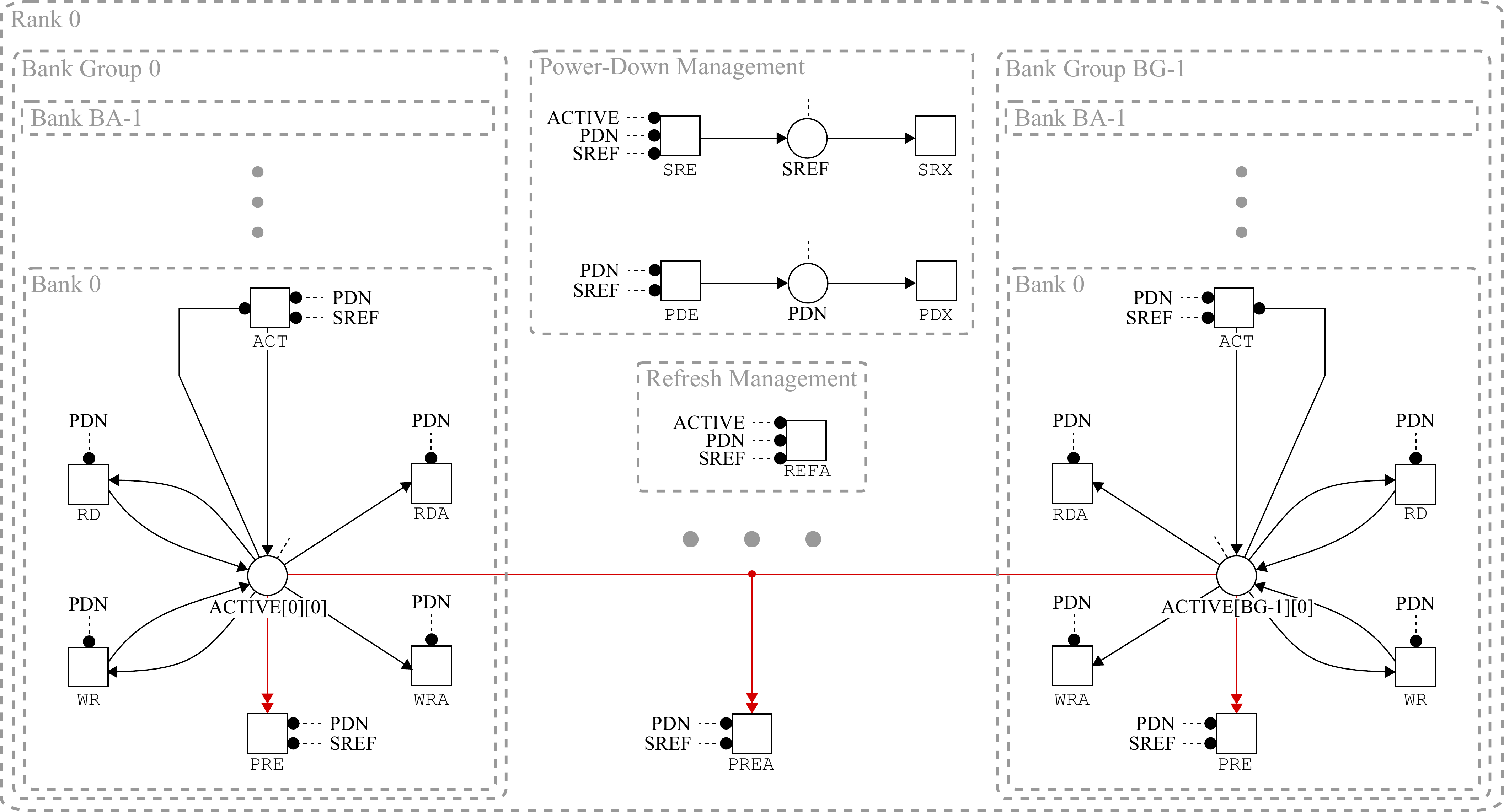}
    \caption{Revised DDR4 Petri Net Model~\cite{junkra_19}}
    \label{fig:petrinet}
\end{figure*}
\subsection{Timed Petri Nets and DRAMml}\label{sec:dramml}
Whenever a new DRAM standard is released by JEDEC, developers of memory controllers and memory simulation models must adopt the changes of the new protocol in order to guarantee compliance of their designs. JEDEC standards describe the memory protocols, i.e., legal DRAM command sequences and timings between commands, using a mixture of state machine diagrams, tables, and timing diagrams. However, even JEDEC themselves admit that parts of the state diagrams are simplified and some events are not captured in full detail~\cite{boxddr412}. Therefore, it is a difficult task for developers to extract the required information from the standards in a comprehensive and formally correct way. To facilitate this task, the authors of~\cite{junkra_19} have introduced a domain-specific language called \textit{DRAMml}, which can serve as a unique formal description of a JEDEC standard's protocol. The language is both compact and extendable, so it can also be used to describe the latest standards with all their additional features in a compressed form. The semantics of DRAMml are based on \textit{timed Petri nets}~\cite{jacjac_11}. 

Because of the large number of parallel operating entities within each channel (e.g., ranks and banks), modeling a current DRAM subsystem with a finite state machines leads to a state explosion. Petri nets, on the other hand, allow the modeling of the DRAM subsystem's concurrency very efficiently.

As shown in Figure~\ref{fig:petrinet}, the hierarchies of a DRAM subsystem are directly adopted in the Petri net. DRAM commands can be mapped to transitions, while different states (i.e., ACTIVE, PDN, SREF) are modeled by the assignment of tokens to places. Constraints in the sequence of commands are modeled by four types of arcs:
\begin{enumerate}
    \item Place-to-transition arcs (\myarcs) allow the connected transition to fire only if a token is present in the connected place. E.g., the arc between an ACTIVE state and a \texttt{RD} command implies that the \texttt{RD} command can only be executed if the respective bank is active. Firing the transition consumes one token.
    \item Transition-to-place arcs (\myarcs) add one token to the connected place if the connected transition fires.
    \item Inhibitor arcs (\myarco) inhibit the connected transition from firing (i.e., command to be executed) if a token is present in the connected place. For example, a \texttt{REFA} command can only be issued if the DRAM is neither in ACTIVE nor in PDN nor in SREF state.
    \item Reset arcs (\myarcd) remove all tokens from the connected place(s) if the connected transition fires. E.g., if a \texttt{PREA} command is executed all tokens are removed from ACTIVE banks, making them idle again.
\end{enumerate}

\noindent Timing constraints between two transitions are modeled by custom command-to-command timing arcs (\myarctj). For example, between the \texttt{ACT} and the \texttt{RD} command there exists a timing arc with the value of $t_{RCD}$. More complex timing constraints like the four activate window\footnote{The four activate window limits the number of issued activate commands within a rolling time window to four.} require a combination of custom timed arcs and places with timed tokens (see~\cite{junkra_19}).

\lstset{language=DRAMml, basicstyle=\ttfamily}
DRAMml is a domain-specific language to describe the complete Petri net structure in a textual way, while the user does not need to understand the underlying Petri net semantics. E.g., the place-to-transition arc between ACTIVE and \texttt{RD} is specified as \lstinline{ACTIVE ->  RD}, the inhibitor arc between ACTIVE and \texttt{REFA} is specified as \lstinline{ACTIVE -o  REFA}, the reset arc between ACTIVE and \texttt{PREA} is specified as \lstinline{ACTIVE ->>  PREA}, and the timing arc between \texttt{ACT} and \texttt{RD} is specified as \lstinline{ACT -<>  RD (tRCD)}. For a more comprehensive explanation of DRAMml we refer to~\cite{junkra_19}.

\subsection{Formal Verification}

Over the last two decades, functional hardware verification using formal methods has reached a high level of maturity. %
The formal tools that are commercially available today are capable of handling designs of realistic size, and they have become standard components in many industrial hardware design and verification flows. %

The approach presented in this paper is based on formal \emph{property checking}. %
A property checker takes as input an RTL description of a hardware design (e.g., in Verilog/SystemVerilog or VHDL), and a set of properties in a property specification language such as \emph{SystemVerilog Assertions} (SVA). %
A property usually has the form of an implication between an \textit{antecedent} (``assumption'') describing an input scenario to be considered, and a \textit{consequent} (``commitment'') describing the behavior expected from the design according to the specification~\cite{2008-NguyenThalmaier.etal}.
%
%
Both assumption and commitment are specified by logic formulas over design signals at different time points. %
Property specification languages such as SVA contain a rich set of syntactic elements that enable the user to write concise and expressive specifications of design intent. %

For each property, the tool searches for a sequence of inputs to the design such that the property is \emph{violated}. %
Such a sequence is called a \emph{counterexample} to the property. %
If the property checker finds a counterexample it is returned to the user as a timing diagram similar to a simulation waveform, displaying all relevant signal values of the RTL model and indicating the time points and signals where the commitment of the property is being violated. %
If the tool reports that no counterexample exists, the property is guaranteed to hold for the design under any possible input sequence. %
It is this mathematical rigor that allows detecting bugs in corner-case states and hard-to-activate input scenarios, which are difficult to anticipate for a human verification engineer. %

In our approach we use formal property checking to verify that a memory controller design fulfills the specification of a particular JEDEC standard expressed in DRAMml. %
Section~\ref{sec:generation} describes how the individual Petri net elements underlying the DRAMml specification are translated into SVA properties that can then be verified for a given controller. %

\section{DDR4 Petri Net}\label{sec:petri_net}
As explained in Section~\ref{sec:dramml} the semantics of DRAMml are based on timed Petri nets. In \cite{junkra_19} the authors have presented an example Petri net for the DDR3 JEDEC standard. However, in this work we aim to verify a memory controller for the successor standard DDR4, which introduces bank groups as an additional device hierarchy. Moreover, the Petri net in \cite{junkra_19} does not fully comply with the specified protocol because it allows a precharge command to be issued only if the associated bank is active. This constraint is not described in the standard. Therefore, we present a revised version of the Petri net in Figure~\ref{fig:petrinet}, which fully complies with the DDR4 standard. It includes bank groups and allows issuing precharge commands also when the respective bank is idle. The key difference in the structure is the elimination of the common idle place. Instead, some additional inhibitor and reset arcs ensure proper command cycling. For the sake of clarity, no timing arcs are shown in the figure.

\section{Property Generation}\label{sec:generation}
The main task of our framework is the generation of SVA from a DRAMml description of a standard by parsing the input file and translating all language elements into SVA equivalents. In the following, the most important translations are explained with short examples. 

As shown in Section~\ref{sec:dramml}, DRAMml describes a timed Petri net with different hierarchies that correspond to ranks, banks, etc. These hierarchies consist of places, transitions and different types of arcs. First, all hierarchies are translated into nested loop generate constructs where the number of iterations is equivalent to the number of instances in each hierarchy (see Listings~\ref{lst:loop_dramml} and \ref{lst:loop_sva}).  

\noindent
\begin{minipage}{\linewidth}
\begin{lstlisting}[
language=DRAMml,
frame=single,
basicstyle=\scriptsize\ttfamily,
caption={DRAMml Hierarchy},
label={lst:loop_dramml}
]
<num_inst> : <hier> {
    // placeholder
}
\end{lstlisting}
\end{minipage}

\noindent
\begin{minipage}{\linewidth}
\begin{lstlisting}[
language=SVA,
frame=single,
basicstyle=\scriptsize\ttfamily,
caption={SVA Hierarchy},
label={lst:loop_sva}
]
genvar <hier>_id;
generate
    for (<hier>_id = 0; <hier>_id < <num_inst>; <hier>_id++) begin
        // placeholder
    end
endgenerate
\end{lstlisting}
\end{minipage}

Each place is translated into a logic variable whose value, i.e., number of tokens, depends on all connected place-to-transition, transition-to-place and reset arcs (see Listings~\ref{lst:place_dramml} and \ref{lst:place_sva}). Modeling a place as a separate logic variable is required to specify properties that consider only input and output signals of a controller design, but do not need access to internal signals. Places with more than one token can be implemented as packed arrays, initial tokens can be assigned during reset. If a place is located inside a hierarchy, an additional dimension is added to the array. Similarly, for transitions within a hierarchy the target instance (i.e., \texttt{<hier>\_id}) has to be considered. 


\noindent
\begin{minipage}{\linewidth}
\begin{lstlisting}[
language=DRAMml,
frame=single,
basicstyle=\scriptsize\ttfamily,
caption={DRAMml Place},
label={lst:place_dramml}
]
Places {
    <place>;
}
Arcs {
    <trans_in> ->  <place>;
    <place>    ->  <trans_out>;
    <place>    ->> <trans_reset>;
}
\end{lstlisting}
\end{minipage}

\noindent
\begin{minipage}{\linewidth}
\begin{lstlisting}[
language=SVA,
frame=single,
upquote=true,
basicstyle=\scriptsize\ttfamily,
caption={SVA Place},
label={lst:place_sva}
]
logic <place>;

always @(posedge clk) begin
    if (reset)
        <place> <= 1'b0;
    else begin
        if (cmd == <trans_in>)
            <place> <= <place> + 1'b1;
        else if (cmd == <trans_out>)
            <place> <= <place> - 1'b1;
        else if (cmd == <trans_reset>)
            <place> <= 1'b0;
    end
end
\end{lstlisting}
\end{minipage}

Self loops, i.e., a place and a transition that are connected by two arcs in a loop (e.g., the \texttt{RD} transition and ACTIVE place within a bank in Figure~\ref{fig:petrinet}), do not have to be added to the always block because they would decrease the variable by 1 and increase it again by 1 in the same clock cycle.

Properties are generated by three different types of arcs. First, a place-to-transition arc is translated into an overlapped implication (see Listings~\ref{lst:arc_assert_dramml} and \ref{lst:arc_assert_sva}). If the command denoted by the transition is executed (assumption), a token must be present in the connected place (commitment). Otherwise, the command sequence is illegal. 

\noindent
\begin{minipage}{\linewidth}
\begin{lstlisting}[
language=DRAMml,
frame=single,
basicstyle=\scriptsize\ttfamily,
caption={DRAMml Place-To-Transition Arc},
label={lst:arc_assert_dramml}
]
Arcs {
    <place> -> <trans>;
}
\end{lstlisting}
\end{minipage}

\noindent
\begin{minipage}{\linewidth}
\begin{lstlisting}[
language=SVA,
frame=single,
basicstyle=\scriptsize\ttfamily,
caption={SVA Place-To-Transition Arc},
label={lst:arc_assert_sva}
]
property arc_<place>_<trans>;
    @(posedge clk) disable iff (reset)
        (cmd == <trans>) |-> (<place> >= 1'b1);
endproperty;

assert property(arc_<place>_<trans>);
\end{lstlisting}
\end{minipage}

An inhibitor arc is translated in a similar way. A place with a token implies the non-firing of the connected transition, i.e., the command is not allowed to be executed in the current state (see Listings~\ref{lst:inhibitor_dramml} and \ref{lst:inhibitor_sva}). 

\noindent
\begin{minipage}{\linewidth}
\begin{lstlisting}[
language=DRAMml,
frame=single,
basicstyle=\scriptsize\ttfamily,
caption={DRAMml Inhibitor Arc},
label={lst:inhibitor_dramml}
]
Arcs {
    <place> -o <trans>;
}
\end{lstlisting}
\end{minipage}

\noindent
\begin{minipage}{\linewidth}
\begin{lstlisting}[
language=SVA,
frame=single,
basicstyle=\scriptsize\ttfamily,
caption={SVA Inhibitor Arc},
label={lst:inhibitor_sva}
]
property inhibitor_<place>_<trans>;
    @(posedge clk) disable iff (reset)
        (<place> >= 1'b1) |-> not (cmd == <trans>);
endproperty;

assert property(@(posedge clk) inhibitor_<place>_<trans>);
\end{lstlisting}
\end{minipage}

The custom command-to-command timing arc is translated into a negated implication with a timing window (see Listings~\ref{lst:timing_dramml} and \ref{lst:timing_sva}), which means that after issuing the command denoted by \texttt{trans\_a} (assumption), a command of the type denoted by \texttt{trans\_c} cannot be issued during the specified timing window (commitment). 

\noindent
\begin{minipage}{\linewidth}
\begin{lstlisting}[
language=DRAMml,
frame=single,
basicstyle=\scriptsize\ttfamily,
caption={DRAMml Command-To-Command Timing Arc},
label={lst:timing_dramml}
]
Arcs {
    <trans_a> -<> <trans_c> (<timing>);
}
\end{lstlisting}
\end{minipage}

\noindent
\begin{minipage}{\linewidth}
\begin{lstlisting}[
language=SVA,
frame=single,
basicstyle=\scriptsize\ttfamily,
caption={SVA Command-To-Command Timing Arc},
label={lst:timing_sva}
]
property timing_<trans_a>_<trans_c>;
    @(posedge clk) disable iff (reset) 
        (cmd == <trans_a>) |-> 
                not ##[1:(<timing> - 1)] (cmd == <trans_c>);
endproperty;

assert property(timing_<trans_a>_<trans_c>);
\end{lstlisting}
\end{minipage}

Depending on the location of arcs and their connected places as well as transitions within the Petri net, the properties are inserted inside the loop generates and the target instances are taken into consideration (e.g., for a \texttt{RD} command the target bank, bank group and rank is also checked). 

More complex timing constraints like the four activate window, which are modeled by a combination of custom timed arcs and places with timed tokens (see Section~\ref{sec:dramml} and \cite{junkra_19}), cannot directly be translated into an implication with a timing window as shown in Listing~\ref{lst:timing_sva}, because they depend on a sequence of commands over a period of time. However, \cite{datsin_08} presents a method to check such constraints with SVA. We use the same approach in our generator and refer to this paper for more information. 



%
%
%
%
%
%
%


\section{Experimental Results}\label{sec:results}
The versatility of our framework is demonstrated by three application examples. First, the standard compliance of an open-source DDR4 controller is verified and its unsupported protocol features are identified. Second, we show how a slightly modified subset of properties can be used to detect performance limitations that result from overestimated timings between commands. Third, we introduce a methodology similar to the idea of property-driven development to upgrade an existing memory controller to a different JEDEC standard.



\subsection{Evaluation of Standard Compliance and Unsupported Features}\label{sec:example_verification}
Due to the lack of open-source controllers for the latest JEDEC standards DDR5, LPDDR5 or HBM3, a low power, low latency DDR4 controller~\cite{controller_chirag,sudlap_19a} is used for demonstration purposes. In order to generate properties for the target hardware, several input parameters of the generator have to be configured. These are the standard, specific timing values in clock cycles and the number of banks, bank groups and ranks. Afterwards, the generated SVA code as well as the RTL description of the controller can directly be read by a property checker, in our case OneSpin~360~DV.

We verify two different controller configurations. In the first case, the default DDR4 configuration with 16 banks distributed over 4 bank groups is used. In the second case, the memory controller is reconfigured to a stripped down version that only supports 8 banks distributed over 2 bank groups. This configuration is required to drive special DDR4 devices with a wider 16 bit data bus. The number of ranks is fixed to 1 because the controller only provides a single chip select signal and, thus, does not support multi-rank operation.

In the case of 16 banks all generated properties hold, which proves that the controller complies with the DRAM protocol. With the reduced number of banks, 4 different properties are violated. Since all violated properties are related to bank group timings, it suggests that the controller was not intended to be operated with only 2 bank groups and this special configuration was not tested. However, with our framework it was straightforward to detect the protocol violations because input parameters are restricted to valid values and both configurations could be checked within one day.

Besides verifying whether a property holds, the checker also proves if the scenarios described in the properties are reachable, i.e., if there exists a sequence of inputs that triggers the corresponding commands. 
For the given DDR4 controller, some properties are in fact unreachable. Even if this does not result in any protocol violations, it means that some parts of the standard have not been implemented, which, in the worst case, prevents the controller from proper operation. Fortunately, the unreachable properties exclusively relate to read and write commands with auto-precharge\footnote{Read and write with auto-precharge allow to trigger a read or write operation and a subsequent precharge with a single command.} and power-down entry and exit commands, which are both optional features.

The possibility to identify unsupported features is extremely helpful for designers that use third party IPs because they can directly compare the feature set of a controller with their requirements without having any prior knowledge and without checking the documentation. A summary of the total number of generated properties, unique properties (i.e., without considering loop generates), unique violated properties and unique unreachable properties for both controller configurations is provided in Table~\ref{tab:assertions}.
\begin{table}
\centering
    \caption{Overview of Properties for DDR4 Controller} 
    \label{tab:assertions}
    \begin{tabular}{|c|c|c|c|c|}
		\hline
		\textbf{Banks} & \textbf{Generated} & \textbf{Unique} & \textbf{Violated} & \textbf{Unreachable} \\
		\hline
		16 & 448 & 124 & 0 & 91\\
        8 & 270 & 124 & 4 & 91\\
		\hline
	\end{tabular}
\end{table}

\begin{figure*}[t]
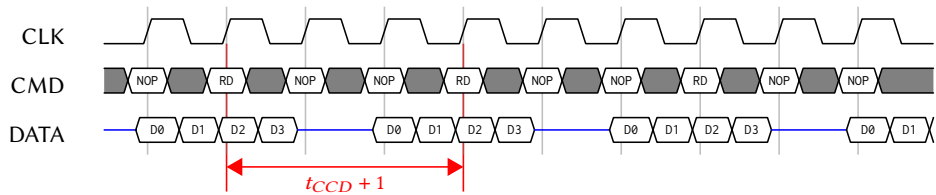

\centering
\begin{tikztimingtable} [timing/d/background/.style={fill=white},
    timing/lslope=0.2, xscale=2.40, yscale=1.5,]
       {\sffamily CLK} & 
       L
       N(P1)
       2{T}
       N(P2)
       2{T}
       N(P3)
       2{T}
       N(P4) 
       2{T} 
       N(P5)
       2{T}
       N(P6)
       2{T} 
       N(P7)
       2{T} 
       N(P8)
       2{T} 
       N(P9)
       2{T} 
       N(P10)
       2.0{T}\\
       {\sffamily CMD} & 
       0.5U O 1U R 1U O 1U O 1U R 1U O 1U O 1U R 1U O 1U O 1.5U\\
       {\sffamily DATA} &
       [black] 0.8Z
       1D{\texttt{D0}}
       1D{\texttt{D1}}
       1D{\texttt{D2}}
       1D{\texttt{D3}}
       2Z
       1D{\texttt{D0}} 
       1D{\texttt{D1}} 
       1D{\texttt{D2}}
       1D{\texttt{D3}}
       2Z
       1D{\texttt{D0}} 
       1D{\texttt{D1}} 
       1D{\texttt{D2}}
       1D{\texttt{D3}}
       2Z
       1D{\texttt{D0}} 
       1D{\texttt{D1}} 
       0.2D{}\\
  \extracode
  \begin{pgfonlayer}{background}

      \timemeasure{P2}{P5}{5.0}{\small $t_{CCD}+1$}

    \begin{scope}[semitransparent,semithick]
      \vertlines[gray]{1.1,3.1,...,19.1}
    \end{scope}
 \end{pgfonlayer}
\end{tikztimingtable}%
\captionof{figure}[Controller-Induced Performance Limitation]{Controller-Induced Performance Limitation}
\label{fig:perf_limit}
\end{figure*}

\subsection{Detection of Performance Limitations}
Besides issuing commands according to the protocol, another main objective of the memory controller is to maximize the performance of the DRAM subsystem. This can be achieved, e.g., by reordering incoming requests to increase the number of consecutive accesses to the same row in a bank (called row hits), which is usually done by a scheduler in the front end. However, the back end can influence the performance as well. If all properties hold, it is ensured that the controller is compliant to a specific JEDEC standard (see Section~\ref{sec:example_verification}). Unfortunately, the timing-related properties only prove that the minimum delay between two commands is always observed. The reverse case, i.e., that two commands are actually issued with the minimum delay if possible, is not ensured. In the worst case, the controller could overestimate timings between successive commands and compromise the overall performance. Figure~\ref{fig:perf_limit} shows an example where the memory controller always issues read commands with a distance of $t_{CCD}+1$, while the minimum timing is $t_{CCD}$. As a result, the data bus is only utilized in two of three clock cycles and the maximum bandwidth is reduced by 33\,\%.

Using formal property checking, we are not able to directly prove this reverse case, but we can at least prove whether there exists one scenario in which the two commands are issued with minimum distance. In addition, if no such scenario is found, a lower bound can be determined successively. For this purpose, the framework generates all reachable timing-related properties from Section~\ref{sec:example_verification} with each timing incremented by one clock cycle. If a property still holds, a possible candidate for a design flaw is found, which can then be investigated in more detail. This process is repeated until the point is reached where all properties are violated. 

For the given DDR4 controller, 9 unique properties still hold with one additional clock cycle added. All these properties include one refresh-related command (either precharge all (\texttt{PREA}), which is usually used before a refresh command to precharge all banks at once, or refresh itself (\texttt{REFA})), which shows that the switch from normal operation to refresh operation is not handled optimally in the controller and extra delays are induced. Only after adding 22 additional clock cycles all properties are violated. However, since refresh commands are issued very rarely, the resulting performance impact is negligible. In fact, it is likely that the extra delays are added on purpose to safely meet all timing constraints before refresh operation is started and save hardware resources by implementing a single clock cycle counter for the worst-case timing assumption.



\subsection{Memory Controller Upgrade}\label{sec:example_upgrade}
Due to the large number of recent JEDEC standard releases, upgrading an existing memory controller to support a new protocol is one of the main development tasks. The continuously-increasing feature set and the resulting hardware complexity makes this task very challenging. As a starting point of the upgrade process it is important for a designer to know which existing parts of the hardware can be adopted, which parts have to be changed and which features have to be added to create a properly functioning controller implementation as fast as possible. 

Based on formal property checking, we present a methodology that identifies the minimum set of changes required to perform such an upgrade. For demonstration, the DDR4 controller from the previous sections should be upgraded to the successor standard DDR5. As a first step, a set of properties for DDR5 is generated with input parameters that match the DDR4 input parameters as closely as possible. In the evaluation in Section~\ref{sec:example_verification} the unsupported controller features have already been identified. All DDR5 properties related to these features are discarded as a next step. The remaining DDR5 properties are then verified by the property checker for the DDR4 controller. 

Afterwards, the necessary upgrades can be derived from the results of the verification. A property that is reachable and still holds means that no further action is required. E.g., the legal and illegal command sequences have not changed from DDR4 to DDR5. Similarly, a subset of the timings is still valid. However, there are also timings that have been newly added. E.g., with DDR5, the minimum timing between two write operations to the same bank group is longer than between two read operations, which was still identical with DDR4. Consequently, the corresponding properties are violated and additional checks have to be implemented in the hardware. Some of the properties are also unreachable. They are related to features that have been newly added with DDR5, e.g., same-bank refresh or a second burst length that can be selected on the fly. For these features the designer must make individual decisions depending on whether they are optional or not and whether they are expected to be required for the controller implementation. The upgrade is completed when all relevant properties hold and are reachable.

\section{Related Work}\label{sec:related}
%
The idea of applying formal property checking to DRAM controllers has already been presented by Datta and Singhal back in 2008~\cite{datsin_08}. They write a set of SVA to formally verify the standard compliance of the Sun OpenSPARC DDR2 controller. The problem is that all properties are extracted manually from the DDR2 JEDEC standard and afterwards translated into SVA code, which takes a lot of time and is prone to human error. In addition, there is no guarantee that all timings and legal command sequences are covered. Similar work is presented in~\cite{kasmic_13} for an LPDDR3 controller.

Kayed et al.~\cite{kayabd_14} address the problem of manual extraction and use a timing diagram tool to translate timing diagrams from a JEDEC standard into a Timing Diagram Markup Language (TDML) based format. From this description they automatically generate SVA. While the generation reduces manual work and thus possible sources of human error, the approach still has a major drawback. Since only some parts of the DRAM protocol are illustrated in timing diagrams, the verification is incomplete and does not prove full standard compliance. Especially the legal and illegal command sequences are not covered at all.



MCXplore, a framework presented in~\cite{haspat_16, haspat_18}, uses a model checker to generate counterexamples for properties with an optimal number of memory requests that serve as test templates for RTL DRAM controllers or DRAM simulators. These test patterns can be used both to validate the memory controller front end (scheduling, etc.) and back end (interactions between memory commands). The approach is design independent and provides quick feedback to a designer, but it also does not cover the entire protocol and does not give any formal guarantees.


In~\cite{sahsha_18}, the authors use the language of Symbolic Analysis Laboratory (SAL) to model the architectural design of a DRAM cache controller in terms of interacting transition systems. This design is then verified using model checking. The work covers lots of cache-related properties, but only includes a subset of protocol constraints. 

The paper~\cite{lisasa_22} proposes a framework written in Coq to model DRAM controllers. Using the framework, the authors prove the legality of command sequences as well as timings between commands, and in addition they check if each incoming request is handled in bounded time. However, as in other works, only a subset of the protocol is checked and a reviewer has to verify that the formal specification matches the standard. 






\section{Conclusion}\label{sec:conclusion}
In this paper we have presented a framework that automatically generates SVA for the formal verification of DRAM controllers. The framework is useful for different tasks of controller development, which was demonstrated in three application examples. During the experiments, we have also detected protocol violations in an open-source DDR4 controller. As a next step, we plan to find more applications of formal verification in memory controller development to easily handle future JEDEC standard releases.


\begin{acks}
This work was supported within the Fraunhofer and DFG cooperation programme (Grant no. 248750294) and supported by the Fraunhofer High Performance Center for Simulation- and Software-based Innovation. Furthermore, we thank Bryan Daniel Olmos Suquillo for his support.
\end{acks}

\bibliographystyle{ACM-Reference-Format}


%
%
%
%
%
%
%
%

\end{document}